\documentclass[a4paper,prf,aps,10pt]{revtex4}
\usepackage{amsfonts,amsmath,amssymb}			
\usepackage{bm}
\usepackage{textgreek}
\usepackage{graphicx}
\usepackage{color}
\usepackage{hyperref}
\hypersetup{
    colorlinks = true,
    citecolor = blue,
    linkcolor = blue,
    filecolor = magenta,      
    urlcolor = green,
}
\usepackage{natbib}
\usepackage{siunitx}									
\usepackage[dvipsnames]{xcolor}

\definecolor{mygray}{gray}{0.5}
\newcommand{\seclab}[1]{\label{sec:#1}}
\newcommand{\secref}[1]{Section~\ref{sec:#1}}
\newcommand{\particles}[1]{#1-$\SImum$-diameter particles}
\newcommand{\figref}[1]{Fig.~\ref{fig:#1}}
\newcommand{\figureref}[1]{Figure~\ref{fig:#1}}
\newcommand{\SImum}{\textrm{\textmu{}m}}
\newcommand{\etal}{\textit{et~al.}}
\newcommand{\SImm}{\textrm{mm}}
\newcommand{\figlab}[1]{\label{fig:#1}}
\begin{document}

\title{Size-dependent particle migration and trapping in 3D microbubble streaming flows}

\author{Andreas Volk$^{\mathrm{(a)}}$,
Massimiliano Rossi$^{\mathrm{(a,b)}}$,
Bhargav Rallabandi$^{\mathrm{(c)}}$,
Christian J. K\"ahler$^{\mathrm{(a)}}$,
Sascha Hilgenfeldt$^{\mathrm{(d)}}$,
Alvaro Marin$^{\mathrm{(e)\dagger}}$}

\affiliation{$^{a}$Institute of Fluid Mechanics and Aerodynamics, Bundeswehr University Munich, Neubiberg, Germany\\
$^{b}$Department of Physics, Technical University of Denmark, DTU Physics Building 309, DK-2800 Kongens Lyngby, Denmark\\
$^{c}$Department of Mechanical Engineering, University of California, Riverside, California 92521, USA\\
$^{d}$Department  of  Mechanical  Science  and  Engineering, University of Illinois,  Urbana-Champaign,  USA\\
$^{e}$Physics of Fluids, University of Twente, Enschede, The Netherlands.}

\begin{abstract}

Acoustically actuated sessile bubbles can be used as a tool to manipulate microparticles, vesicles and cells. 
In this work, using acoustically actuated sessile semi-cylindrical microbubbles, we demonstrate experimentally that finite-sized microparticles undergo size-sensitive migration and trapping towards specific spatial positions in three dimensions with high reproducibility.
The particle trajectories are successfully \textcolor{black}{reproduced by passive advection of the particles in a steady three-dimensional streaming flow field augmented with volume exclusion from the confining boundaries.}
For different particle sizes, this \textcolor{black}{volume exclusion mechanism} leads to three regimes of qualitatively different migratory behavior, suggesting applications for separating, trapping, and sorting of particles in three dimensions.

\end{abstract}

\maketitle

\section{Introduction}
\seclab{introduction}
The rectification of oscillatory fluid motion into powerful steady flows has been known and understood since the 19th century \cite{Rayleigh1883}. In recent years, due to the developments in manipulation of microparticles, micro-organisms and other microscopic objects \cite{ashkin1970acceleration}, the research on acoustic rectification, streaming and trapping has been strongly revitalized \cite{bruus2011_0}. 
Leveraging the dynamics of high-frequency oscillations and the non-linearity of fluid dynamics, both fluid elements and particles are forced on time-averaged steady trajectories, resulting in {\em streaming flow} and {\em particle migration}, respectively. 
A particularly well-developed way of obtaining vigorous rectification effects is driving a microbubble's interface at ultrasound frequencies. The resulting net flow, known generically as steady streaming \cite{Riley2001}, or, in this particular case, \emph{microbubble streaming}, will advect solid particles placed in the flow volume.  However, the particles will not necessarily behave as passive tracers, but will typically migrate onto specific trajectories and positions, and/or become \emph{entrapped} in certain areas in the bubble's vicinity. 

One of the first observations of such an entrapment was by Miller, Nyborg and Whitcomb \cite{Miller:1979wg}. They observed how human platelets formed aggregates around gas-filled micropores when the liquid was exposed to ultrasound. In a follow-up work, Miller \cite{Miller:1988vr} concluded that attractive acoustic radiation forces generated from the bubble's oscillation could not explain the experimental observations.
Nonetheless, their discovery opened the door for new methods to trap, suspend, manipulate and even shear particles or cells of any kind, including motile cells.
In recent years, with the development of microfabrication technologies and controlled microfluidic experiments, Lutz \emph{et al.} \cite{Lutz:2006gv} and Lieu \emph{et al.} \cite{lieu2012hydrodynamic} generated microvortices -- or \emph{microeddies} -- around solid microcylinders by imposing low-frequency liquid oscillations, showing the ability to trap sufficiently large particles.
That feature gives hydrodynamic manipulation through microbubble oscillations an advantage against competing micro-trapping techniques as optical trapping \cite{ashkin1970acceleration} or dielectrophoretic trapping \cite{wang1997dielectrophoretic}. 

\textcolor{black}{Particle aggregation in vortical flows carrying heavy and/or light particles has been extensively studied in turbulent flows, where tracers can have long residence times in vortices \cite{biferale2005particle,calzavarini2008dimensionality}. 
At lower Reynolds numbers, particles have been observed to aggregate dynamically in preferred orbits inside thermocapillary-driven liquid bridges \cite{schwabe1996new,Kuhlmann2011accumulation,melnikov2013synchronization}, strongly influenced by the boundaries of the liquid bridge. 
Closer to our system, but in the absence of a free surface, is the particle trapping observed in flow eddies generated by steady acoustic streaming close to indentations in a microchannel \cite{Schwartz2012trapping}.}

Nonetheless, the physical mechanism behind the particle trapping phenomenon by oscillating bubbles remains unclear. 
Although the low-frequency trapping results of Lutz \emph{et al.} \cite{Lutz:2006gv} and Lieu \emph{et al.}\cite{lieu2012hydrodynamic} showed trapping was possible even when acoustic radiation was negligible, subsequent work by Rogers and Neild \cite{Rogers:2011bo} and Chen \emph{et al.} \cite{Chen:2016fu} proposed to model such trapping around larger oscillating bubbles using concepts of radiation force. Other authors \cite{chong2016} have focused on a more direct numerical computation of inertial effects due to the finite size and density contrast of the particles in the streaming flow. In recent work, Agarwal \emph{et al.} \cite{AgarwalPRF2018} presented an analytical treatment that showed how particle attraction to or repulsion from an oscillating bubble can indeed be described as a consequence of inertial effects dependent on the driving conditions as well as the particle and fluid properties. Acoustic radiation forces were incorporated into this formalism as a special case.

The controlled experiments in the abovementioned publications were almost exclusively designed to confine particle motion to a plane for easier quantification. However,
Marin \emph{et al.} \cite{marin2015} recently revealed the strong three-dimensional character of such streaming flows precisely due to the confinement in the third dimension. The acoustically-driven streaming flow around semi-cylindrical bubbles analyzed in that work was theoretically described by Rallabandi \emph{et al.} \cite{rallabandi2015} using an asymptotic model reproducing the 3D particle trajectories. Although Marin \emph{et al.}'s experiments were performed using 2-$\SImum$-diameter particles, considering the good comparison with the computed streamlines, it could be concluded that such small particles followed the streamlines faithfully and could be considered as passive flow tracers.

The aim of the current work is to examine and describe the size-sensitive migration and trapping of particles in such a 3D bubble streaming flow, and to understand how to make use of the 3D structure to separate particles along qualitatively different paths.
Due to the intrinsic three-dimensional character of the particle motion, classical two-dimensional measurements methods will fail to resolve the real particle trajectories. Instead, in this work we obtain three-dimensional particle trajectories using a defocusing-based particle tracking method (General Defocusing Particle Tracking) \cite{barnkob2016}.

Using experiments and numerical simulations, we show that the three-dimensional character of the flow field is crucial to understand and exploit the systematic migration and trapping of density-matched particles. Our numerical simulations reveal that the migration and trapping can be well reproduced making use of the 3D stream functions \cite{rallabandi2015} and \textcolor{black}{volume exclusion} constraints with the confining boundaries due to the particle's finite-size.

\section{Experimental Setup}
\seclab{exp_setup}

A sketch of the experimental setup is shown in \figref{expSetup}. The experiments were carried out in a Polydimethylsiloxane (PDMS) microchannel with a rectangular cross-section of $500\,\SImum \times 72 \,\SImum$ and a blind side channel with width $w=90\,\SImum$ and depth $l=350\,\SImum$. The microchannel was fabricated using soft lithography \cite{xia1998} and bonded to a glass slide previously covered with a 1-$\SImm$-thick PDMS film in order to have all channel walls made of PDMS. The bonding was realized by functionalizing the surfaces with a corona plasma treater (Elveflow, France). Once the microchannel is filled with liquid (a water/glycerol mixture, see below), a bubble remains at the blind side channel forming a gas-liquid interface that protrudes into the main channel. The system can be \emph{short-circuited} to stop the flow by connecting the inlet to the outlet with a 4-way valve (see \figref{expSetup}) and the bubble size can be adjusted by controlling the static pressure of the liquid as described earlier \cite{volk2015, volk2018_2}. The experiments in this work were performed with a bubble of width $110\,\SImum$ protruding $50\,\SImum$ into the main channel. Note that as a consequence of the manufacturing process, the edges of the blind side pit are blunt resulting in a slightly wider bubble than the width of the blind side channel (see inset in \figref{expSetup}).

\begin{figure}[t!]
  \centering
  \includegraphics[width=.5\columnwidth]{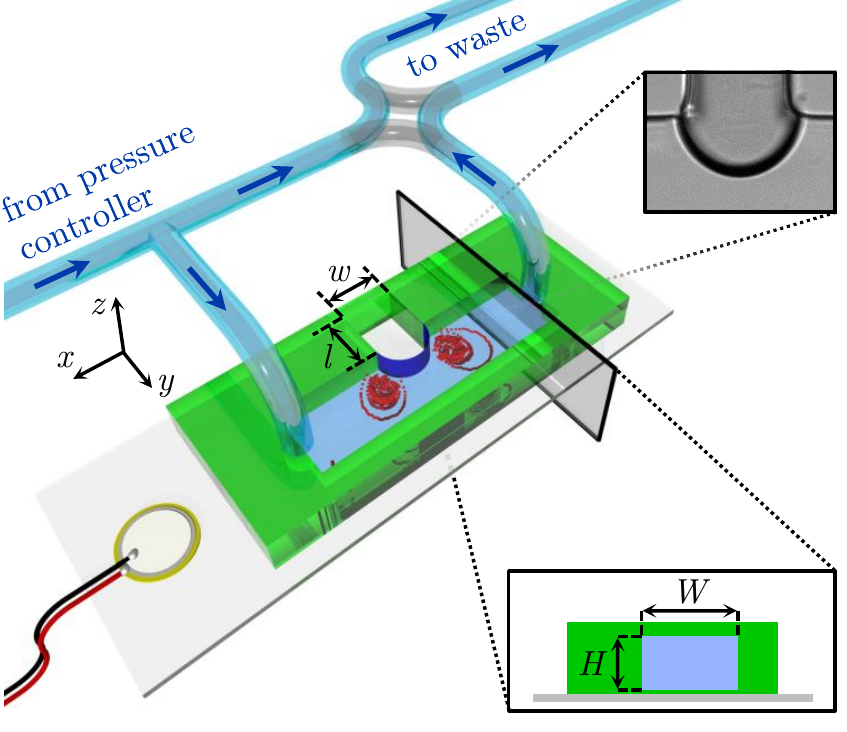}
\caption{\figlab{expSetup} Schematic (not to scale) experimental setup: A PDMS microchannel with width $W=500\,\SImum$, height $H=72\,\SImum$, and total length  $\sim 30\,\mathrm{mm}$ is filled with liquid and a sessile cylindrical bubble is formed in the blind side channel of width $w=90\,\SImum$ and depth $l=350\,\SImum$. By switching the valve (indicated in grey), inlet and outlet of the microchannel are connected, leading to a uniform pressure in the system that stops the flow. The size of the bubble is controlled by adjusting the static pressure of the liquid with a pressure controller. A piezo transducer attached to the glass slide excites surface oscillations at the bubble interface generating a vortical acoustic streaming flow (in red) in the microchannel.}
\end{figure}

The bubble oscillations are driven by a piezoelectric transducer of $1\,\SImm$ thickness and a diameter of  $10\,\SImm$ (Physik Instrumente, Germany), attached to the glass slide with epoxy two-component glue. The structure and magnitude of the streaming flow generated by the bubble surface oscillation depends on the driving frequency and amplitude and the bubble size \cite{wang2013,volk2018_2}. For our investigation, we used an actuation frequency $f=21.8\,\mathrm{kHz}$, a resonant frequency of bubble oscillation, with a peak to peak voltage $U_{pp}=20\,\mathrm{V}$. The streaming flow generated with these settings was used to investigate the behavior of particles of different sizes dispersed in the fluid. In particular, we used fluorescent polystyrene spheres of diameter 2, 5, 10 and $15\,\SImum$ (Microparticles GmbH, Germany) suspended in an aqueous glycerol solution with a glycerol content of 22.6$\%$ w-w to match the particle density \cite{volk2018_1}.

\begin{figure}[t!]
  \centering
  \includegraphics[width=\columnwidth]{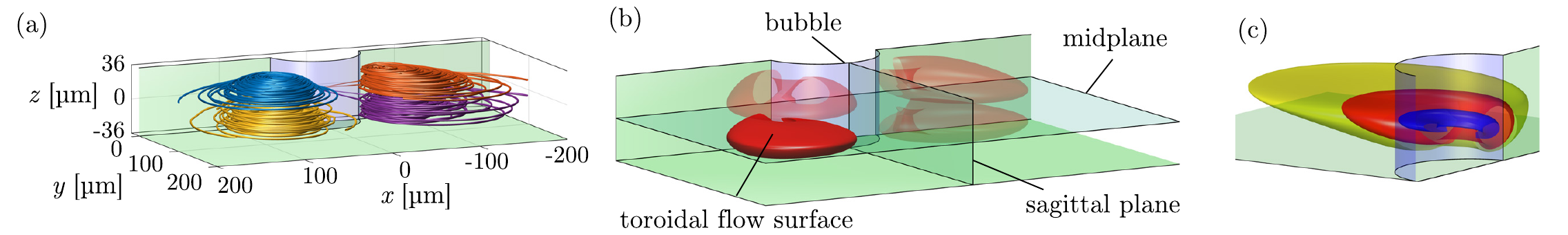}
\caption{\label{fig:flowTorii} Streaming flow structures near an oscillating semi-cylindrical microbubble. (a) Trajectories of 2-$\SImum$-diameter particles. The plot shows one trajectory in each of the equivalent quadrants. (b) Sketch of the flow structures and symmetry planes as observed in experimental investigations. The flow field is set up in four toroidal flow structures. These structures are symmetric to the midplane $z=0$ (spanning in the $xy$-direction) and the sagittal plane at $x=0$ (spanning in the $yz$-direction). (c) A reconstruction of three toroidal flow surfaces reveals the nested structure of these tori.}

\end{figure}

The three dimensional particle trajectories were determined with General Defocusing Particle Tracking (GDPT) \cite{barnkob2016,RuneMassi2020}, a single-camera particle tracking method that uses pattern recognition to identify the depth position of spherical particles from the shape of their corresponding out-of-focus images. Astigmatic optics, consisting of a cylindrical lens placed in front of the camera sensor, was used to enhance the particle image deformation and extend the measurement volume \cite{cierpka2010,rossi2014}. A high speed camera (PCO Dimax, Germany) was used to image the particles in fluorescent mode at frame rates around 300 fps and maximum exposure time. 
For this setup, we used a microscope objective of magnification $10$$\times$ in combination with a cylindrical lens of focal length $500\,\SImm$, yielding a total measurement volume of approximately $2000\,\SImum \times\,2000\,\SImum\times\,100\,\SImum$, and an estimated uncertainty on the particle position of $\sim\,0.5\,\SImum$ in the $xy$-direction and $\sim\,1\,\SImum$ in the $z$-direction \cite{RuneMassi2020}. 

\section{Results}
\subsection{Flow structure around the bubble - Experiments}
\seclab{TrajIntr}

The flow structure around the bubble was investigated using tracer particles with a diameter of $2\,\SImum$, which can be considered as almost-passive flow tracers. \figref{flowTorii} (a) shows the measured trajectories of four selected tracer particles which approximately follow  streamlines of the fluid flow. In general, each particle trajectory is confined to one of the four regions defined by the two symmetry planes of the system:  the midplane $z=0$ and the sagittal plane $x=0$. Away from the bubble, the particles move in almost planar trajectories ($z\approx$ const). But as they approach the bubble, the particles strongly accelerate and acquire a rapid motion along the $z$-axis, effectively ``jumping" to a different $z$-position, with the direction and magnitude of the jumps depending on the loop size and position. Large loops pass closer to the bubble surface and their jumps move particles toward the midplane. Due to the symmetry of the flow, the loops also change along the bubble axis $z$: as the loops approach the midplane, they become tighter and more distant from the bubble surface. At a certain distance from the midplane, the drift in the $z$-direction is reversed, i.e., the particle appears to be reflected at the midplane and then travels towards the channel's top/bottom wall. A similar trajectory reflection occurs as a particle approaches the solid top/bottom walls. This pattern repeats itself over and over, with the trajectories lying on a torus-shaped surface as shown in  \figref{flowTorii} (b). 
The trajectory of each tracer lies on the surface of a unique torus defined by its initial position. Tori of different sizes have a nested structure (as Russian nested dolls), with a small torus contained inside a larger torus as shown in \figref{flowTorii} (c). The innermost torus covers the smallest range of $z$ and has negligible thickness, representing a single, non-planar trajectory, which we will call the {\em coreline} occupying a mean $z$-position of $Z_{core}\approx \pm 18~\SImum$. The plane occupied by the {\em coreline} divides the toroidal structure in two identical halves. More details on these structures were revealed in previous work with experiments and simulations \cite{marin2015, rallabandi2015}. 

Such nested structures can be accessed by fluid elements or passive tracers. In the next section we show how particles are restricted to access certain regions due to their finite size and describe how their trajectories deviate from the fluid pathlines described here.

\seclab{results}

\subsection{Trajectories of differently sized particles - Experiments}
\seclab{TrajExp}
\begin{figure}[ht!]
  \centering
\includegraphics[width=\columnwidth]{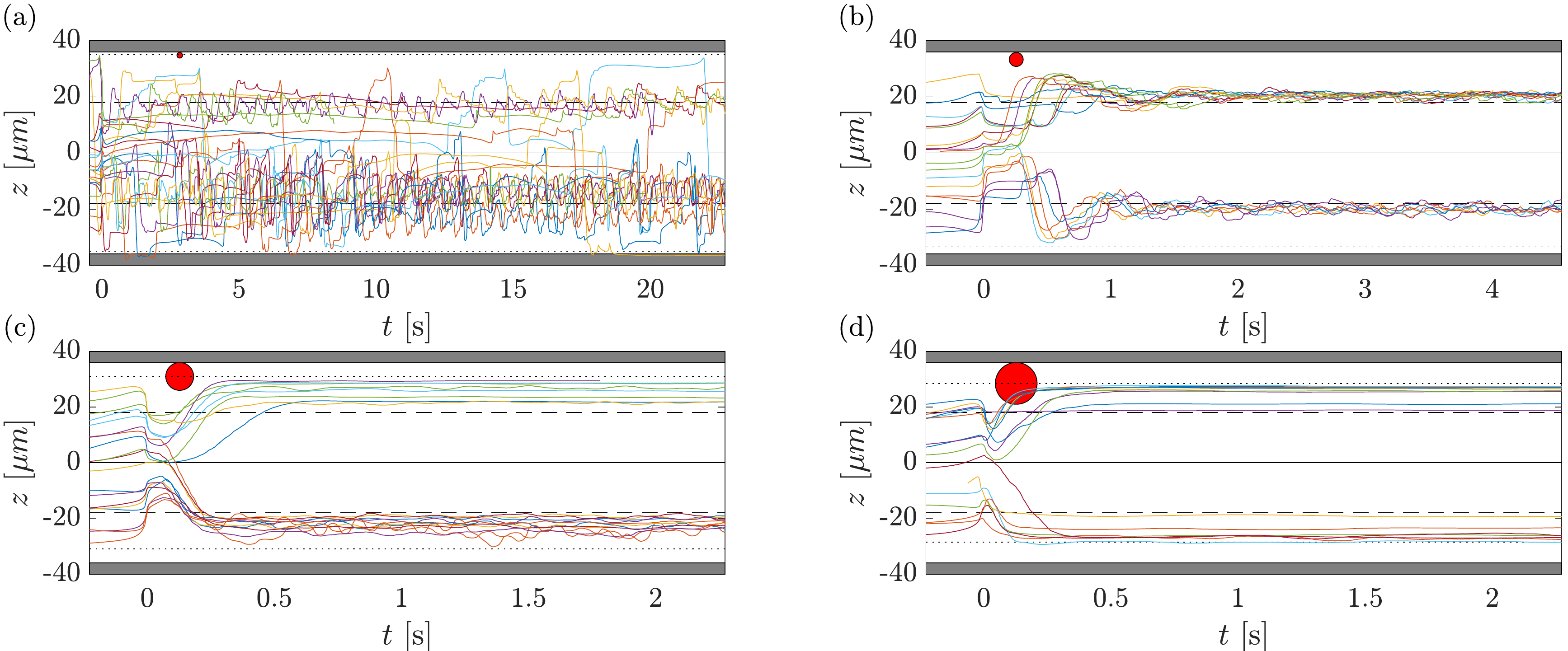}
\caption{\label{fig:ZTsizesExp} Experimentally determined $z$-position of particles with $d_p=$ 2, 5, 10, and 15 $\SImum$ in microbubble streaming flow as a function of time. After the particles pass the perigee for the first time (closest position to the bubble's interface, set to $t=$ 0), they tend to migrate to a certain $z$-position at different rates depending on their size. Dashed lines depict the coreline position $Z_{core}= \pm 18\,\SImum$. A red circle illustrates the size of the particle and dotted lines are drawn at one particle radius from the top and bottom walls. (a) $z$-position for particles with $d_p=$ 2 $\SImum$, they  slowly migrate towards inner tori of the flow structure. Focusing at a defined $z$-position is not observed.  (b) $z$-position for particles with $d_p=$ 5 $\SImum$; they migrate within $\sim$1 second towards $Z_{core}= \pm 18\,\SImum$. (c) particles with $d_p=$ 10-$\SImum$ migrate to different final $z$-position within $\sim$0.3 seconds. (d) particles with $d_p=$ 15 $\SImum$ migrate within $\sim$0.3 seconds and focus at the top and bottom walls of the microchannel. Dashed line: midplane; dotted lines: closest position of the particle center to top and bottom walls.}
\end{figure}

Having described the structure of the streaming flow, we now turn our attention to the trajectories of particles with $d_p=$ 2, 5, 10 and 15 $\SImum$, followed over a time scale of 3 to 30 seconds. Each experiment starts with particles at rest distributed randomly over the whole channel volume. When the ultrasound is turned on and the streaming flow is generated, particles are set into motion. Figure \ref{fig:ZTsizesExp} shows the $z$-position of differently sized particles as a function of time. As can be seen in \figref{ZTsizesExp}a, 2-$\SImum$-diameter particles show a broad distribution within the $z$-range of the channel throughout the whole experiment. As mentioned in \secref{TrajIntr} these particles behave as passive tracers following the streamlines along certain toroidal flow surface given by the particle initial position. The length and amplitude of the periodic \emph{bouncing} in $z$ therefore depends on the specific toroidal flow surface. 
However, a slight narrowing of the collective $z$-range can be observed after an initial transient; particles thus migrate towards inner flow surfaces, i.e., tori of smaller extent in $z$ closer to the coreline at $Z_{core}\approx \pm 18\,\SImum$  (see \figref{ZTsizesExp}a). 
This effect is stronger as particle size increases: particles with $d_p=$ {5} $\SImum$ show a faster and more pronounced migration (\figref{ZTsizesExp}b). When these particles pass the perigee (location closest to the bubble surface), they get displaced towards the eye of a toroidal flow surface and within $\approx 1\,$s the particle is stabilized near $Z_{core}$.

Although the migration process is similar at first glance for larger particles, the mean final $z$-position deviates from $Z_{core}$ if the particle diameter is $d_p \gtrsim 10\,\SImum$: larger particles migrate to positions closer to the top or bottom walls. That is the case for the larger particle sizes explored: $d_p=$ 10 $\SImum$ (\figref{ZTsizesExp}c) and 15 $\SImum$ (\figref{ZTsizesExp}d). 
The swift migration and trapping of large particles is very clearly shown for \particles{15} which, after migrating to the eye of the toroidal flow vortex, end up in contact with either the top or bottom channel wall.
Looking back at \particles{10}, particles in this size-range represent a transition between a regime in which  trajectories tend to gravitate around the coreline (\particles{5}, \figref{ZTsizesExp}b) and a regime where trajectories nearly touch the channel walls (\particles{15}, \figref{ZTsizesExp}d).
In order to get a deeper understanding of the processes that lead to the observed migration and trapping, the particle trajectories will be theoretically modeled and computed in the following section. 

\subsection{Trajectories of differently sized particles - Numerical model} 
\seclab{TrajSim}
\begin{figure}[b]
  \centering
\includegraphics[width=0.33\columnwidth]{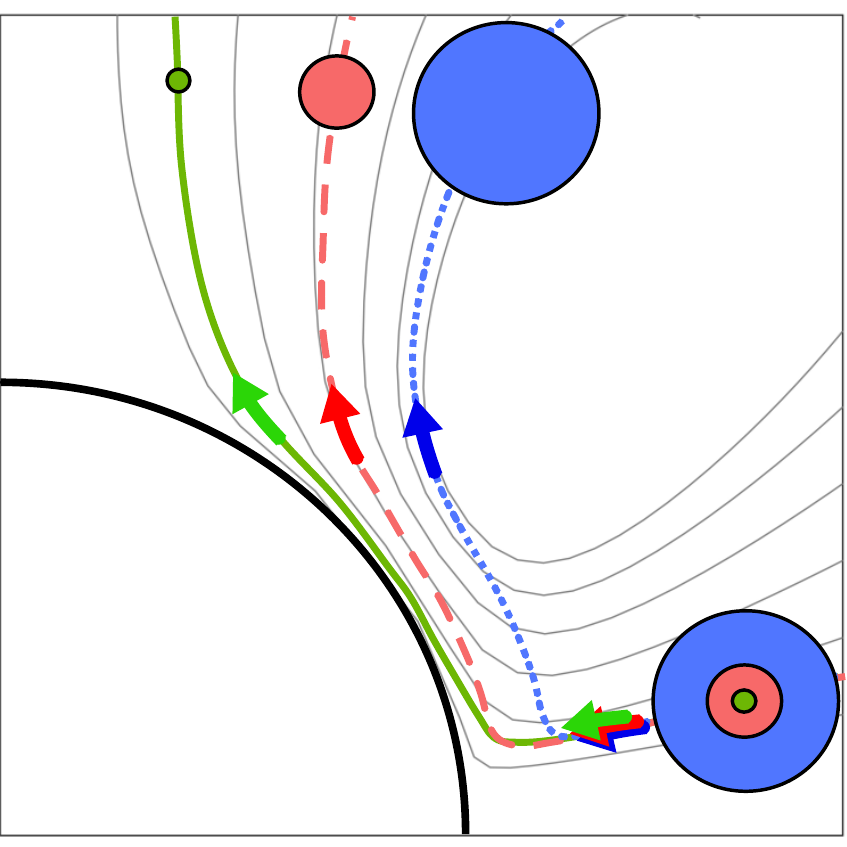}
\caption{\label{fig:SimScheme} Schematic of the \textcolor{black}{excluded volume interaction} introduced in the numerical model for trajectories of differently sized particles. While being advected in the streaming flow of the oscillating bubble, particles approach the bubble (black line) along streamlines (grey lines). A particle that would penetrate the bubble in a certain timestep of the simulation is radially displaced and forced onto another streamline.}
\end{figure}

\begin{figure}[t!]
  \centering
\includegraphics[width=\columnwidth]{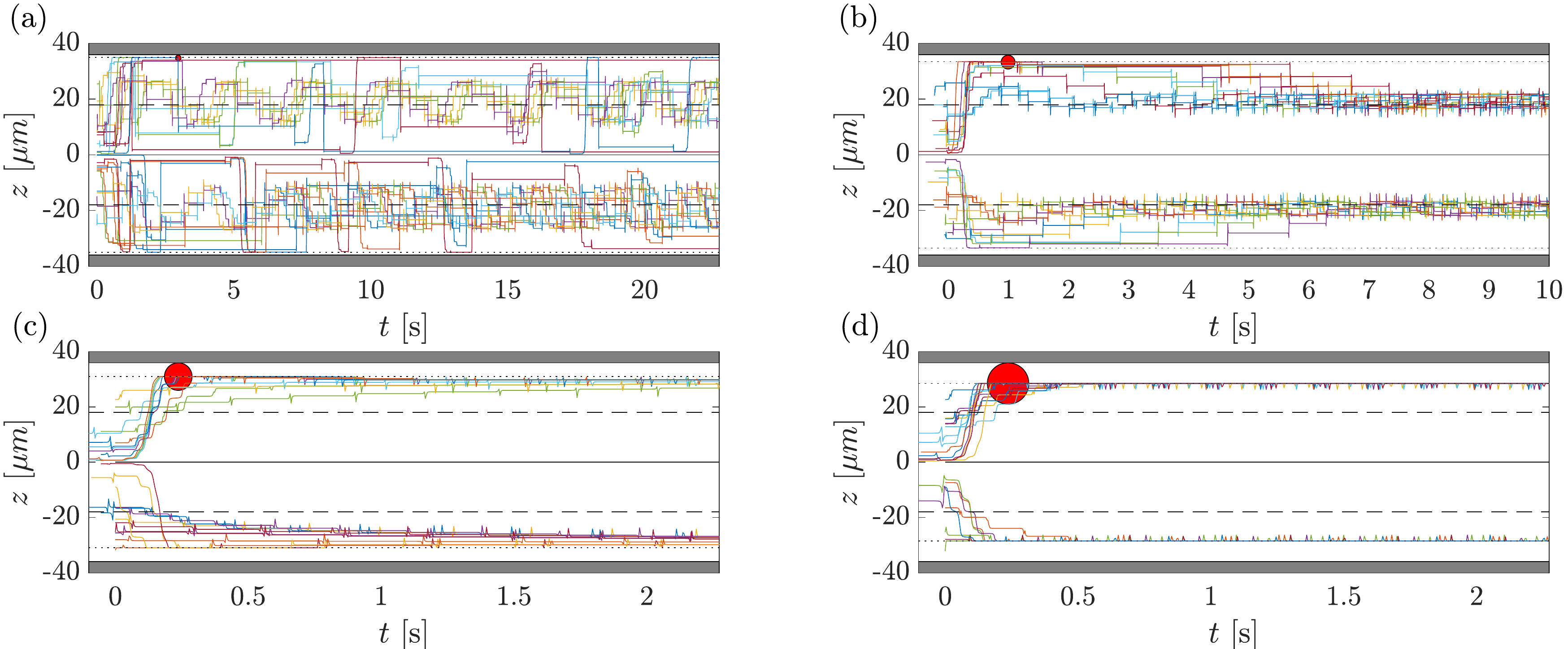}
\caption{\label{fig:ZTsizesSim} Numerical $z$-positions of particles with $d_p=$ 2, 5, 10, and 15 $\SImum$ in microbubble streaming flow as a function of time. After the particles pass the perigee for the first time (closest position to the bubble's interface, set to $t=$0), larger particles start to migrate to a certain $z$-position. Dashed lines depict the coreline position $Z_{core} = \pm 18\,\SImum$. A red circle illustrates the size of the particle and dotted lines are drawn at one particle radius from the top and bottom walls. (a) $z$-positions for particles with $d_p=$ 2 $\SImum$, some slowly migrate towards the inner tori of the flow structure, others span the whole volume of the microchannel, but none focus at a particular $z$-position. (b)Particles with $d_p=$ 5 $\SImum$ clearly migrate towards $Z_{core}$. While the migration times obtained from simulations are comparable to the experimental ones (Fig. \ref{fig:ZTsizesExp}) for most of the particle sizes, the migration of 5-$\SImum$-diameter particles is somewhat slower and therefore a longer time-axis is shown in comparison to Fig. \ref{fig:ZTsizesExp}b. (c) Particles with $d_p=$ 10 $\SImum$ quickly migrate in the vicinity of the microchannel walls. (d) Particles with $d_p=$ 15 $\SImum$ quickly migrate towards the microchannel walls, with frustrated attempts to penetrate the wall.}
\end{figure}

As shown by Rallabandi \etal~\cite{rallabandi2015}, the complex flow structure described above can be reproduced by a superposition of three streaming flow functions. Each function or mode satisfies the appropriate boundary conditions at one of three boundaries: the stress-free bubble surface and the no-slip top/bottom and side walls, respectively. Rallabandi \etal~\cite{rallabandi2015} and Marin \etal~\cite{marin2015} showed that flow tracers follow the streamlines computed from such stream functions.

While the physical nature of inertial forces near oscillating bubbles has been elucidated recently \cite{thameem2017,AgarwalPRF2018}, these computations are elaborate and have only been rigorously developed for 1D and 2D flow trajectories. Fortunately, a simplified scheme has been developed before \cite{Wang:2012hg,thameem2016}, in which the inertial forces are replaced by \textcolor{black}{simple volume exclusion}: When the trajectory of  a finite-size particle approaches the bubble to within a particle radius, the particle position is displaced in a direction normal to the bubble surface such that particle-interface interpenetration is prevented over the next numerical integration step (see \figref{SimScheme}). This principle is easily extended to our 3D flows, applying the same type of displacement whenever the particle touches a side wall or the top/bottom wall of the channel. 
We acknowledge that a proper description of particle motion in a flow like this involves more sophisticated modeling of rectified forces. However the dominant effect on particles close to interfaces has been found to be lubrication repulsion \cite{thameem2017}, which, due to its short range and strong magnitude, can be replaced in good approximation by a \textcolor{black}{excluded volume interaction}. 

The numerical model has been tested by introducing particles of different sizes at different initial positions in a systematic way. Time is normalized in the simulation by making use of the circular frequency of the oscillation $\omega$ and the bubble's oscillation amplitude relative to its radius, which we denote as $\epsilon$. In the simulations, time is cast in a dimensionless variable $\tilde{t} = \epsilon^2 \omega t$, where $(\epsilon^2 \omega)^{-1}$ represents the characteristic timescale for the streaming flow. The choice of $\epsilon$ determines the maximum steady flow velocity $|v_\mathrm{{max}}| = \epsilon^2 \omega a_b$, so that a value of $\epsilon = 0.05$ yields velocities of order of the experimentally observed 20 mm/s. 
The results are shown in \figref{ZTsizesSim}, where one can observe a clear migration of finite-sized particles when the particle diameter is larger than $2\,\SImum$ (\figref{ZTsizesSim}a). For the numerical time range explored, \textcolor{black}{small \particles{2} faithfully follow the streamlines, covering the wide toroidal flow surface,} whereas larger \particles{5} (\figref{ZTsizesSim}b) get displaced to stream-surfaces closer to the coreline of the toroidal flow structure. When the particle diameter is increased to 10 $\SImum$ (\figref{ZTsizesSim}c) and 15 $\SImum$ (\figref{ZTsizesSim}d), particles typically get trapped in the vicinity of the top/bottom walls. 
Although the computed detailed particle trajectories differ from those in experiment, the particle dynamics in Fig. \ref{fig:ZTsizesSim} compares favorably in time scale and final position for all particle sizes explored. 

\begin{figure}[b]
  \centering
\includegraphics[width=.5\columnwidth]{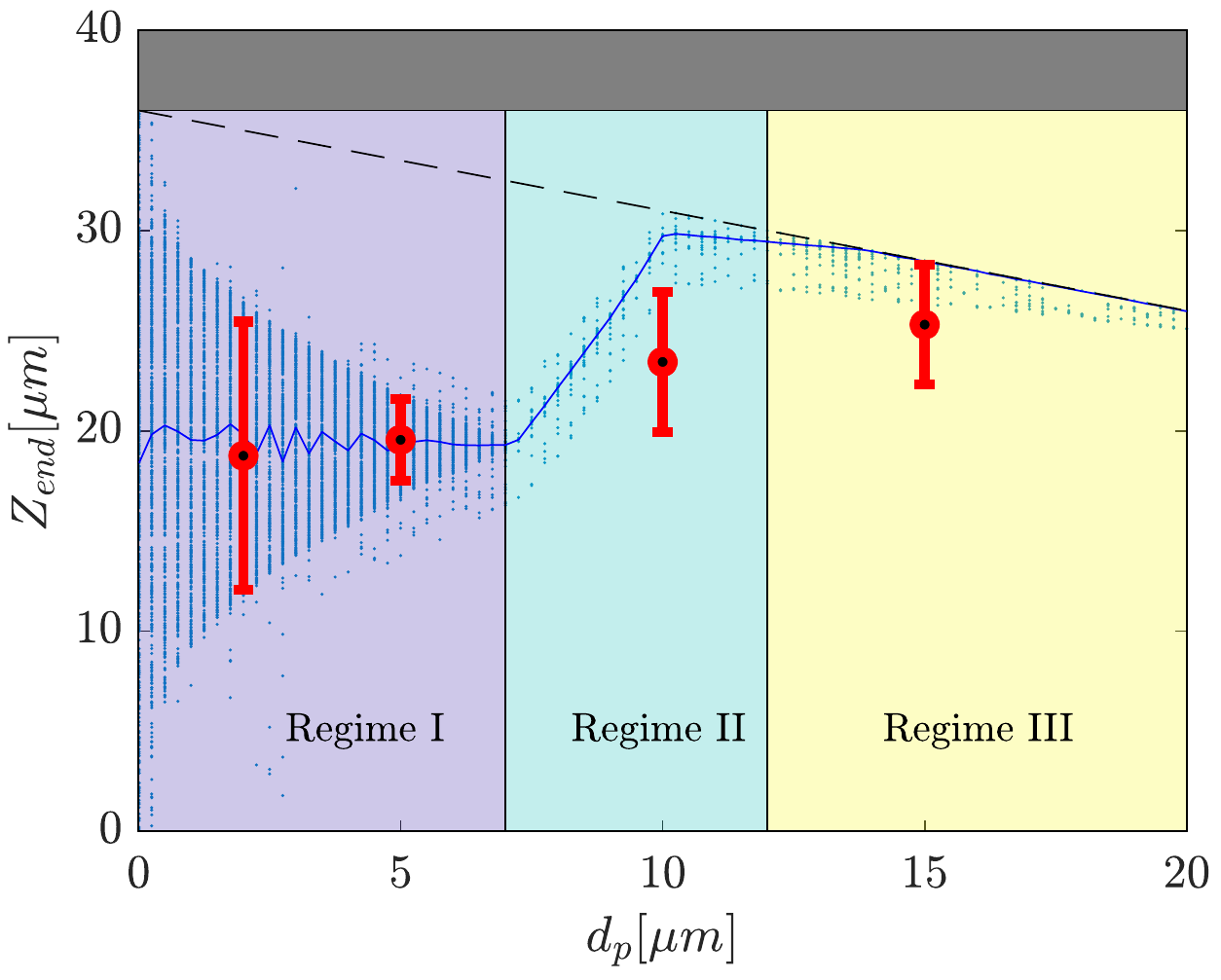}

\caption{\label{fig:ZendCompare} Final particle positions $Z_{end}$ as a function of particle diameter $d_p$. Black circle markers show the average $Z_{end}$ for experimental trajectories of a certain particle size and the red bar is the respective standard deviation. Small blue dots show $Z_{end}$ for many simulated particle trajectories and the blue line the corresponding mean. The PDMS channel is indicated in gray with an exclusion zone that particles cannot enter due to their finite size (dotted black line). Due to the channel symmetry experimental results are mirrored to positive values at the midplane ($z=0$), while simulations use only initial particle positions with $z>0$.}
\end{figure}

In order to make a more quantitative comparison between experiments and simulations, we compute the final $z$ position of particles $Z_{end}$ of different sizes in the range 0.1 $\mathrm{\mu m}$ to 20 $\mathrm{\mu m}$, with initial positions homogeneously distributed along the whole numerical volume. 
{The final position $Z_{end}$ is defined as the last position of the particle in a long simulation or in a long experiment. 
Since the particle initial positions are homogeneously distributed in a systematic way in simulations, and randomly in the experiments, a resulting narrow distribution in $Z_{end}$ for a certain particles size $d_p$ is indicative of a strong trapping.
Both simulations and experiments are run for a dimensionless time period of $\Delta \tilde{t} \approx $ 3000 for all particles (about 10 seconds for typical experimental conditions), except for \particles{2} and below, which are run for $\Delta \tilde{t} > $ 10000 (more than 30 seconds in experiment). The longer time periods approach the maximum storage capability of the high-speed camera; time periods in simulations are matched to these experimental conditions.}

\begin{figure}[t!]
  \centering
\includegraphics[width=\columnwidth]{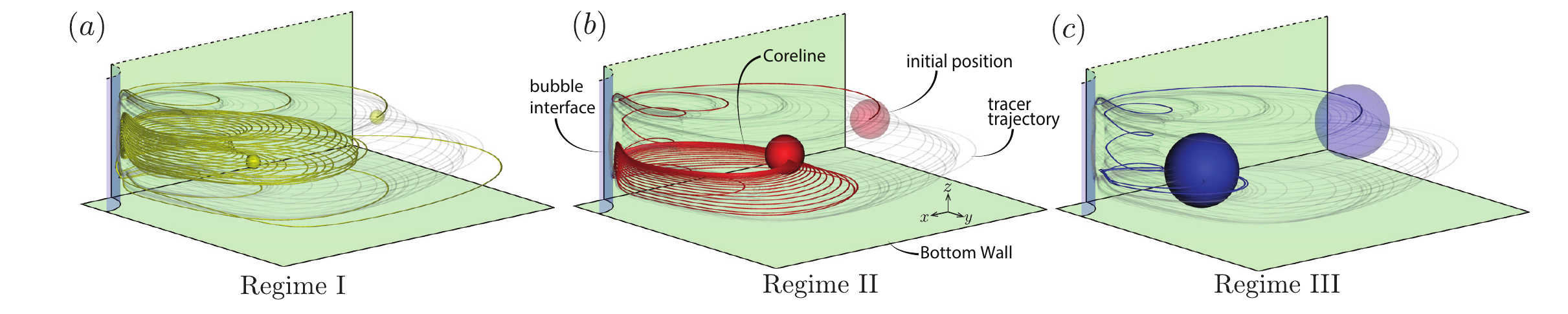}
\caption{\label{fig:typicalTracks} Typical particle trajectories computed from simulations in the three regimes introduced in \figref{ZendCompare} for identical initial conditions. \textcolor{black}{Particles follow their computed streamline until displaced by excluded volume interactions}, as described in \secref{TrajSim}. A tracer particle trajectory with the same initial condition (defined as a particle with $d_p=0~\SImum$) is plotted in semi-transparent gray. Semi-transparent spheres indicate the particle's initial position and opaque spheres indicate the last particle position computed in the simulation. (a) particle with $d_p=$ {3} $\SImum$ showing an initial displacement and then passive advection on a  flow torus. (b) \particles{8} follow a part of a torus, cut short by interactions with the bubble near the coreline. (c) Typical trajectory of \particles{15}, which quickly migrate towards the bottom wall and remain trapped there.}

\end{figure}

The computed final positions $Z_{end}$ are plotted as a function of the particle size and shown in \figref{ZendCompare}.
The experimental results are shown as the mean of $Z_{end}$ and the error bar represents its standard deviation (24 trajectories for particles with $d_p=$ {2} $\SImum$, 19 for {5} $\SImum$, 19 for {10} $\SImum$ and 16 for {15} $\SImum$).
{The ensemble of trajectories for each particle size yields robust values of the mean and the standard deviation (Fig. S1 in supplementary material) with negligible sensitivity on the choice of the final position selected for obtaining $Z_{end}$.}
The spread of $Z_{end}$ values for the smallest particle size range in both experiments and simulations reflects the uniform initial distribution of particles throughout the fluid volume. \figureref{ZendCompare} shows a good match of the computed final particle location with those found in the experiments. It also shows three different regimes for different particle sizes that were already anticipated while analyzing the experimental results. Typical trajectories in such three regimes are also depicted in Fig. \ref{fig:typicalTracks} and will be discussed in the following section. 
\section{Discussion: size-dependent regimes}

{Three regimes are identified according to the type of migration and/or trapping experienced and shown in the trend revealed by $Z_{end}$ (Fig. \ref{fig:ZendCompare}) for the different particle sizes and initial conditions. We proceed to describe each regime in more detail in the following lines.}

{The smallest particles in regime I do not migrate to a specific $z$ plane but remain on toroidal flow surfaces (see \figref{typicalTracks}a), following them passively according to their initial condition. Finite particle size prevents them, however, from occupying the outermost tori, on which the particle approaches the bubble so closely as to intersect with it.  Volume exclusion then moves the particle to an inner torus, which is then followed passively. Thus, the smallest particles have the widest range of $Z_{end}$. 
This range decreases for increasing particle size until the particles are forced to occupy the coreline, i.e., the innermost torus at $Z_{end}\approx 18\,\SImum$ (see \figref{ZendCompare}), with the smallest span in $z$. Such a sharp focusing is already visible in the experimental data for 5-$\SImum$-diameter particles and it is clearly seen in the simulations. 
The limiting particle size for reaching the coreline is $d_p\approx 7\,\SImum$, where regime I ends.}

{As the particle diameter is increased further above approximately $ 7\,\SImum$ (see \figref{ZendCompare}), particles cannot avoid interactions with the bubble, even on the coreline. Consequently, \textcolor{black}{volume exclusion} by the bubble surface now takes place every time the bubble is approached (radial distance in the $xy$-plane is small enough). By this ``kick'', the particle is forced into the eye of the toroidal structure inside the coreline, which is part of a larger torus and thus moves the particle towards the top/bottom wall of the microchannel \footnote{Note that the system is symmetrical with respect to the channel midplane, and therefore the same discussion applies to the top and bottom walls.}.}
{As shown in Fig. \ref{fig:typicalTracks}b, the particle is passively advected on the larger torus, eventually reversing $z$-motion back towards the mid-plane in wider loops. Due to the large size of the particles, they must again interact with the bubble and are kicked back into the eye of the tori. Thus, in regime II large particles move along a portion of wider tori (see yellow torus in Fig. \ref{fig:flowTorii}c) but with their $z$-range is restricted to a region near the top/bottom wall due to the bubble's presence ``cutting short'' the completion of the toroidal  pattern.}
{The precise $d_p$ of the onset of regime II in experiment is not known due to the limited particle size range available. But the migration effect seems to be somewhat weaker in experiments for approximately \particles{10} than in the model; nonetheless, the transition (I$\rightarrow$II) is clearly represented in both the experimental data and the model.
}

{As the particle size increases further, a new constraint comes into play: when particles are transported towards the top/bottom wall, the turning point of reversal of $z$-motion on the torus they are following is closer to the wall than a particle radius. At this point, \textcolor{black}{volume exclusion repulsion} with the wall is inevitable, hindering further motion towards the wall. As the flow tries to advect the particle towards the wall, it becomes trapped on an essentially planar trajectory in almost continuous contact with the wall. This \textcolor{black}{excluded volume interaction} with the wall characterizes the new regime III (Fig. \ref{fig:typicalTracks}c).  Above $d_p\approx 10~\SImum$ all particles in the simulations and most particles in the experiments are trapped in this way.}

{In summary, in regime I, particles may experience a one-time interaction with the bubble, which limits the amount of tori available; but they are able to explore a complete flow-torus in a stable, passive trajectory permanently. In regime II, no complete tori are accessible to the particles; they are only allowed to explore part of a torus, whose extent is defined by the \textcolor{black}{excluded volume} interaction with the bubble happening once per cycle. In regime III, even the partial torus cannot be followed, as the particle trajectory is constrained by interaction with one of the channel walls, yielding a near-planar confinement at the wall.}

\section{Discussion: Limitations}

{We have shown that, by simply advecting finite-sized particles through \emph{a priori} known solutions for the steady streaming flow, one can obtain good quantitative agreement for the final migration position between experimental data and numerical results. 
However, a few significant disagreements have been found: (1) \particles{2} migrate towards their final trajectories on somewhat slower transients than is observed in the simulations. (2) The transitional regime II (\figref{ZendCompare}) seems less pronounced in experiments than in simulations. (3) The ending position for larger particles (regime III in \figref{ZendCompare}) is broader in experiments than in simulations.}

Several phenomena could account for such discrepancies. Probably the most important effect that has not been considered so far is the presence of inertial forces. Recent experiments with oscillating bubbles have shown the presence of rectified lift forces on particles \cite{thameem2016,thameem2017,AgarwalPRF2018}. Inertial forces have ranges beyond the \textcolor{black}{excluded volume repulsion} introduced here and thus may explain the greater migratory tendency of \particles{2} in experiments (Fig. \ref{fig:ZTsizesExp}) versus simulations (Fig. \ref{fig:ZTsizesSim}). It is however not the aim of this work to study in detail the role of the inertial forces in the long term migration of these particles. 
{The effect of inertial forces is also enhanced by large flow velocity. In this study we have worked with maximum particle velocities (close to the bubble) in the range of 20 mm/s (compatible with bubble amplitude of oscillations in the range of $\epsilon=0.05$). A stronger actuation of the bubble might increase the role of inertial forces in the particle dynamics.}

{Although the particle concentration employed in each of the experiments is extremely low (less than 5 particles per image and per experiment), particle-particle interactions might occur occasionally. This is more likely to occur during the experiments with larger particles in regime III (\figref{ZendCompare}), where two or three large particles could have perturbed each other, which might explain the wider distributions of positions in experiments in contrast with the simulations.}

Another effect of crucial importance that has not been considered in this work is the role of particle-liquid density mismatch. {All our results have been performed with polystyrene particles in density-matched liquid solutions with errors below 1\% in density \cite{volk2018_1}.} Trajectories of particles with larger and smaller density ratios could be altered by buoyancy as well as additional inertial lift forces \cite{AgarwalPRF2018}.

\textcolor{black}{While the boundaries of the regimes shown in Fig. \ref{fig:ZendCompare} were obtained for the specific streaming flow determined by the bubble geometry, size, and the driving frequency, it was shown that the topology of bubble microstreaming flows is very robust against changes of these parameters \cite{rallabandi2014}. The exact location of regime boundaries may be parameter dependent, but the existence of regimes I-III only depends on this flow topology and the presence of the confining boundaries interacting with the finite-size particles.}

\section{Conclusions and future work}

We have shown that the \textcolor{black}{final position} of a particle introduced into a streaming flow can only be understood when taking into account the 3D flow structure, even in a case designed to emphasize 2D flow in a certain plane of the microchannel. Our quantitative experimental measurements and numerical simulations demonstrate that the eventual position of particles in the bubble's axial direction depends strongly and predictably on the particle size. While small enough particles behave as passive tracers and follow the streamlines \cite{marin2015, rallabandi2015}, larger particles migrate consistently, not only in the $xy$-plane, but also along the $z$-axis onto certain limiting trajectories in a very reproducible manner. The timescale of this migration depends strongly on the particle size and on the flow velocity.

Although a more careful study should include proper modeling of the inertial rectification forces known to act on the particles \cite{thameem2017,AgarwalPRF2018,chong2016}, \textcolor{black}{ our results show that the size-selective and long-term migration of the particles is physically determined by the interplay of advection with the flow and short-range repulsion from boundaries, here modeled by a simple volume exclusion mechanism.} \textcolor{black}{The main ingredients of this theory are the finite size of the particles, and their interaction with the boundaries. This concept is applicable to many boundary-driven streaming flows, beyond the specific case of bubble microstreaming. Indeed, similar interactions with boundaries have been used in the context of thermal Marangoni flows in liquid bridges \cite{schwabe1996new} to explain the particle migration to and accumulation on specific trajectories \cite{Kuhlmann2011accumulation}.}

{The three-dimensional particle migration revealed in our work is a very attractive option in microfluidic devices for particle trapping and size-sensitive sorting applications \cite{Wang:2011ip,Wang:2012hg}. Our simple model can serve as a useful tool to design flow geometries in which particles of different sizes can be separated and recovered. This should work most efficiently in the most confined regimes (namely II and III), in which particles will separate in $z$-position according to their size. Once they have reached their $Z_{end}$ position, the acoustic actuation can be stopped and a directional flow along the channel switched on to collect them in different positions. Note that the approach provides particle size sorting by $z$-position in addition to the already established separation in $y$ \cite{Wang:2011ip,thameem2017} in a setup providing throughput in $x$. This could be of particular interest for constricted microfluidic systems in which solid particles are prone to clog the flow due to sieving of large particles or arching \cite{Dressaire:2017bc,marin2018clogging,marin2020clogging}.
Future work will be dedicated to understand the role of inertial forces in the long-term migration observed in these systems and to analyze the role of density mismatch. 

\section*{Acknowledgements}
This work is the culmination of many years of work and collaboration between these authors, and many others not included in the author list have contributed indirectly to it. At least we should acknowledge abundant discussions and technical help from Raqeeb Thameem, Cheng Wang and Rune Barnkob. AV and CJK acknowledge financial support by the German Research Foundation grant KA 1808/17-1. AM acknowledges the financial support by the European Research Council via the Starting Grant number 678573. 

\bibliographystyle{unsrtnat}

\end{document}